# OB Associations, Wolf-Rayet Stars, and the Origin of Galactic Cosmic Rays

W.R. Binns[*], M.E. Wiedenbeck[†], M. Arnould[‡], A.C. Cummings[**],
G.A. de Nolfo[***], S. Goriely[‡], M.H. Israel[*], R.A. Leske[**], R.A. Mewaldt[**],
G. Meynet[††], L. M. Scott[*], E.C. Stone[**], and T.T. von Rosenvinge[***]

*Washington University, St. Louis, MO 63130 USA*

[†]*Jet Propulsion Laboratory, California Institute of Technology, Pasadena, CA 91109 USA*

[‡]*Institut d'Astronomie et d'Astrophysique, U.L.B., Bruxelles, Belgique*

[**]*California Institute of Technology, Pasadena, CA 91125 USA*

[††]*Geneva Observatory, 1290 Sauverny, Switzerland*

[***]*NASA/Goddard Space Flight Center, Greenbelt, MD 20771 USA*

(*Author for correspondence: Email: wrb@wuphys.wustl.edu)

**Abstract.** We have measured the isotopic abundances of neon and a number of other species in the galactic cosmic rays (GCRs) using the Cosmic Ray Isotope Spectrometer (CRIS) aboard the ACE spacecraft. Our data are compared to recent results from two-component (Wolf-Rayet material plus solar-like mixtures) Wolf-Rayet (WR) models. The three largest deviations of galactic cosmic ray isotope ratios from solar-system ratios predicted by these models, $^{12}C/^{16}O$, $^{22}Ne/^{20}Ne$, and $^{58}Fe/^{56}Fe$, are very close to those observed. All of the isotopic ratios that we have measured are consistent with a GCR source consisting of ~20% of WR material mixed with ~80% material with solar-system composition. Since WR stars are evolutionary products of OB stars, and most OB stars exist in OB associations that form superbubbles, the good agreement of our data with WR models suggests that OB associations within superbubbles are the likely source of at least a substantial fraction of GCRs. In previous work it has been shown that the primary $^{59}Ni$ (which decays only by electron-capture) in GCRs has decayed, indicating a time interval between nucleosynthesis and acceleration of $>10^5$ yr. It has been suggested that in the OB association environment, ejecta from supernovae might be accelerated by the high velocity WR winds on a time scale that is short compared to the half-life of $^{59}Ni$. Thus the $^{59}Ni$ might not have time to decay and this would cast doubt upon the OB association origin of cosmic rays. In this paper we suggest a scenario that should allow much of the $^{59}Ni$ to decay in the OB association environment and conclude that the hypothesis of the OB association origin of cosmic rays appears to be viable.

# 1. Introduction

Previous observations have shown that the $^{22}Ne/^{20}Ne$ ratio at the GCR source is greater than that in the solar wind (e.g. Wiedenbeck & Greiner 1981; Mewaldt et al. 1980;

Lukasiak et al. 1994; Connell & Simpson 1997; DuVernois et al. 1996). Several models have been proposed to explain the large $^{22}$Ne/$^{20}$Ne ratio (Woosley & Weaver 1981, Reeves 1978, Olive & Schramm 1982, Cassé & Paul 1982, Prantzos et al. 1987, Maeder & Meynet 1993, Soutoul & Legrain 1999, and Higdon & Lingenfelter 2003). (See Binns et al. (2005) and Mewaldt (1989) for a more detailed discussion of these models). Cassé and Paul (1982) first suggested that ejecta from Wolf-Rayet stars, mixed with material of solar system composition, could explain the large $^{22}$Ne/$^{20}$Ne ratio. Prantzos et al. (1987) later developed this idea in greater detail. The WC phase of WR stars is characterized by the enrichment of the WR winds by He-burning products, especially carbon and oxygen (Maeder & Meynet 1993). In the early part of the He-burning phase, $^{22}$Ne is greatly enhanced as a result of $^{14}$N destruction through the α-capture reactions $^{14}$N(α,γ)$^{18}$F(e$^+$,ν)$^{18}$O(α,γ)$^{22}$Ne (e.g. Prantzos et al. 1986; Maeder & Meynet 1993). A high elemental Ne/He ratio in the winds of WC stars has been observed (Dessart et al. 2000), which is consistent with a large $^{22}$Ne excess. The high velocity winds from WR stars can inject the surface material into regions where standing shocks, formed by those winds and the winds of the hot, young, precursor OB stars interacting with the interstellar medium (ISM), may pre-accelerate the WR material.

Kafatos et al. (1981) originally suggested that cosmic rays might be accelerated in superbubbles. Streitmatter et al. (1985) showed that the observed energy spectra and anisotropy of cosmic rays were consistent with such a model. Streitmatter & Jones (2005) have recently shown that the first and second "knees" above ~$10^{15}$ and $10^{17}$ eV in the all-particle energy spectrum may be explained in the context of a superbubble model. A model in which particles might be accelerated to energies above $10^{18}$ eV by multiple SN explosions in OB associations was developed by Bykov & Toptygin (2001 and references therein). Parizot et al. (2004) further explored the collective effects of shocks within superbubbles on cosmic ray acceleration. Higdon and Lingenfelter (2003) have argued that GCRs originate in superbubbles based on the excess of $^{22}$Ne/$^{20}$Ne in GCRs. In earlier work, they pointed out that most core-collapse supernovae (SNe) and WR stars occur within superbubbles (Higdon et al. 1998). In their model, WR star ejecta and ejecta from core-collapse SNe within superbubbles mix with interstellar material of solar-system composition, and that material is accelerated by subsequent SN shocks within the superbubble to provide most of the GCRs. Higdon & Lingenfelter (2003) conclude that the elevated $^{22}$Ne/$^{20}$Ne ratio is a natural consequence of the superbubble origin of GCRs since most WR stars exist in OB associations within superbubbles.

## 2. Measurements

The CRIS instrument consists of four stacks of silicon detectors for dE/dx and total energy ($E_{tot}$) measurements, and a scintillating-fiber hodoscope that measures particle trajectories (Stone et al. 1998). The dE/dx-$E_{tot}$ method is used to determine particle charge and mass. The CRIS geometrical factor is ~250 cm$^2$sr and the total vertical thickness of silicon available for stopping particles is 4.2 cm. The angular precision that is obtained by the fiber hodoscope is ≤ 0.1°.

The neon data used in this paper were collected from 1997 Dec. 5 through 1999 Sept. 24 and are a high-resolution, selected data set. Events were selected with trajectory angles ≤25° relative to the detector normal, and particles stopping within 750 μm of the dead layer surface of each silicon wafer were excluded from the analysis. Nuclei interacting in CRIS were rejected using the bottom silicon anticoincidence detector, which identifies penetrating particles, and by requiring consistency in the multiple mass estimates that we calculate. Additionally, particles with trajectories that exit through the side of a silicon stack were also rejected (Binns et al. 2005).

The average mass resolution for neon that we obtained for energies over the range of ~85 to 275 MeV/nucleon is 0.15 amu (rms). This is sufficiently good that there is very little overlap of the particle distributions for adjacent masses, as shown in Figure 1. The total number of neon events is ~4.6 × 10$^4$. Histograms of F and O isotopes that are used in the GCR propagation model to obtain the $^{22}$Ne/$^{20}$Ne ratio at the cosmic ray source are also shown in this figure.

## 3. Source Composition

The $^{22}$Ne/$^{20}$Ne abundance ratio at the cosmic-ray source is obtained from the ratio observed using a "tracer method" (Stone & Wiedenbeck 1979), which utilizes observed abundances of isotopes that are almost entirely produced by interstellar interactions of primary cosmic rays to infer the secondary contribution to isotopes like $^{22}$Ne, for which the observed fluxes are a mixture of primary and secondary nuclei. The isotopes that we have used as tracers are $^{21}$Ne, $^{19}$F, and $^{17}$O. The cross-sections used in the propagation model are described in Binns et al. (2005), where details of the model can be found.

Combining the results obtained using these three tracer isotopes, Binns et al. (2005) obtained a "best estimate" of the $^{22}$Ne/$^{20}$Ne ratio of 0.387 ± 0.007 (stat.) ± 0.022 (syst.).

Expressing this as a ratio relative to solar wind abundances (Geiss, 1973), we obtain $(^{22}Ne/^{20}Ne)_{GCRS}/(^{22}Ne/^{20}Ne)_{SW}$ ratio of $5.3 \pm 0.3$.

## 4. Wolf-Rayet Model Comparison

Supernovae (SNe) shocks are believed to be the accelerators of GCRs up to energies of at least $\sim 10^{15}$ eV. Most core-collapse supernovae (SNe) in our galaxy (~80-90%) are believed to occur in OB associations within superbubbles (Higdon & Lingenfelter 2003; 2005). Furthermore, most WR stars are located in OB associations and most of the O and B stars with initial mass $\geq 40$ M$_\odot$ are believed to evolve into WR stars. These massive stars have short lifetimes, e.g. ~4 million years for a 60 M$_\odot$ initial mass star, and their WR phase lasts for typically a few hundred thousand years (Meynet & Maeder 2003; Meynet et al. 1997). The most massive stars with the shortest lifetimes evolve through their WR phase injecting WR wind material, including large amounts of $^{22}$Ne, into the local circumstellar medium, which is already a low density bubble resulting from coalescing main-sequence star winds (Parizot, et al. 2004; van Marle et al. 2005). The shocks from SNe in the OB association should sweep up and accelerate both their own ejected pre-supernova wind material and WR wind material from the more massive stars in the OB association.

The mass of the neon isotopes synthesized by massive stars in their WR and core-collapse SN phases and ejected into superbubbles has been estimated by Higdon and Lingenfelter

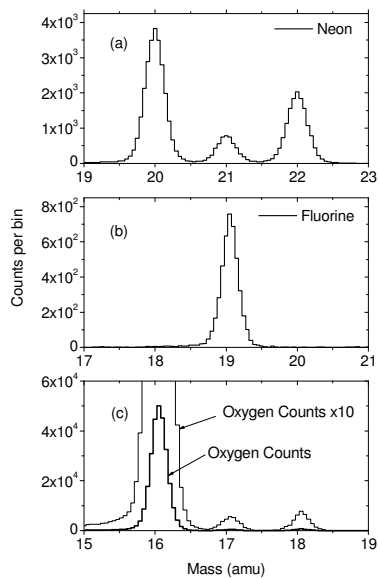

(2003). Based on these calculations, they estimate that a mass fraction, 18±5%, of WR plus SN ejecta, mixed with ISM material of solar-system composition results in the $^{22}$Ne/$^{20}$Ne ratio reported in an earlier analysis of the CRIS results (Binns et al. 2001). They state that "the $^{22}$Ne abundance in the GCRs is not anomalous but is a natural consequence of the superbubble origin of GCRs in which the bulk of GCRs are accelerated by SN shocks in the high-metallicity WR wind and SN ejecta enriched interiors of superbubbles".

Figure 1— Mass histograms for (a) neon, (b) fluorine, and (c) oxygen. The neon energy range extends from ~85 to 275 MeV/nucleon. The solar modulation parameter derived for these observations is $\phi$=400±60 MV (Binns et al. 2005) which gives a midpoint energy for neon in the local ISM of ~380 MeV/nucleon.

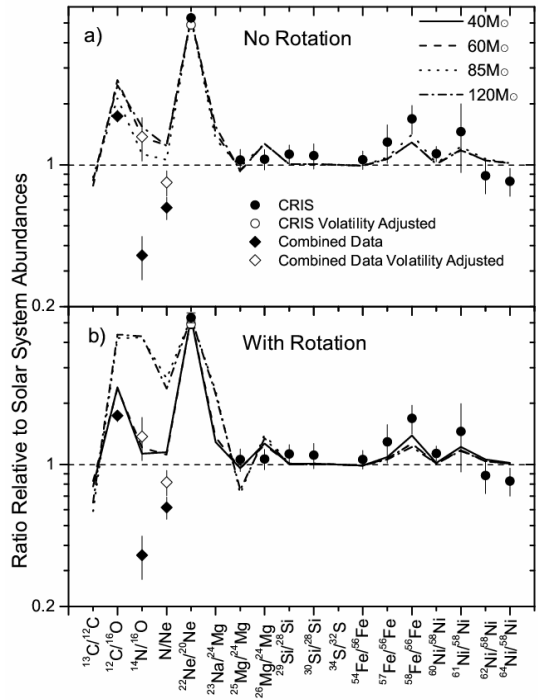

We have examined other isotope ratios at the cosmic-ray source, inferred from our CRIS observations and other experiments, as an additional test of the origin of cosmic rays in OB associations. In Figure 2 we compare these ratios with two-component modeling calculations of WR outflow (Binns et al., 2005; Meynet & Maeder 2003) for metallicity Z=0.02 and initial precursor-star rotational equatorial-velocities at the stellar surface of either 0 or 300 km/s. For each WR star model we calculated the mixture (by mass) of WR wind material with material of solar-system (solar-wind) composition required to give the CRIS $^{22}$Ne/$^{20}$Ne GCR source ratio. Table 1 shows the mass fraction of the total cosmic ray source material required from the WR star since its initial formation for each case.

Figure 2—CRIS ratios compared with model predictions for WR stars with (a) no rotation, and (b) an equatorial surface rotation velocity of 300 km s$^{-1}$ for the initial precursor star for masses of 40, 60, 85, and 120 M$_\odot$, and for metallicity Z$_\odot$=0.02.

Table 1—The mass fraction of ejecta from WR stars, integrated from the time of star formation, mixed with ISM material of solar-system composition, that is required to normalize each model to the CRIS $^{22}$Ne/$^{20}$Ne ratio.

| WR Initial Mass (M$_\odot$) | No-rotation WR Fraction | Rotating WR Fraction |
|---|---|---|
| 40 | -- | 0.22 |
| 60 | 0.20 | 0.16 |
| 85 | 0.12 | 0.44 |
| 120 | 0.16 | 0.37 |

Although the material being mixed with ISM material in the two approaches (Binns et al., 2005, and Higdon & Lingenfelter, 2003) is slightly different (i.e. Higdon & Lingenfelter include neon contributions from SNe in the material being mixed with ISM), the mixing fractions are very similar with the exception of the higher fractions predicted for the very rare M ≥ 85 M$_\odot$ rotating stars.

In each of the two-component models described above, the material ejected from massive stars is mixed with ISM with an assumed solar system composition to normalize to the

$^{22}$Ne/$^{20}$Ne composition. It might be questioned whether ISM with solar system composition is the right material to mix with the ejecta from massive stars. However, Reddy et al. (2003) show that although the abundances of heavy elements increase slowly with time as the galaxy evolves, the ratio of pairs of heavy elements change by only small factors (Also see Wiedenbeck et al. 2007). Thus, since we are examining ratios of heavy isotopes and elements, it would appear that the use of ISM with solar system abundances is a reasonable approximation to reality.

The CRIS results are plotted as closed circles in Figure 2 (see Wiedenbeck et al. 2001a,b and 2003 for elements heavier than neon). Ulysses Mg and Si data (not plotted; Connell & Simpson 1997) are in good agreement with our CRIS results, but their $^{58}$Fe/$^{56}$Fe ratio is significantly lower than the CRIS value (Connell 2001). Wiedenbeck et al. 2001b discuss a possible reason for this. The lighter elements are plotted as solid diamonds and are mean values of GCR source abundances, relative to solar system, obtained from Ulysses (Connell & Simpson 1997), ISEE-3 (Krombel & Wiedenbeck 1988; Wiedenbeck & Greiner 1981), Voyager (Lukasiak et al. 1994) and HEAO-C2 (Engelmann et al. 1990). (See Binns et al. 2005 for more details.) The error bars are based on weighted means from these experiments. The solar-system abundances of Lodders (2003) are used to obtain the abundances relative to solar system.

In Figure 2 we see that, for nuclei heavier than neon, the WR models agree well with the data, except for the high-mass (85 and 120 solar masses) rotating-star models that show a deficiency in the $^{25}$Mg/$^{24}$Mg ratio, which is not observed in GCRs. In particular, the observed enhancement of $^{58}$Fe/$^{56}$Fe is consistent with the model predictions.

For elements lighter than neon there is usually only a single isotope for which source abundances can be obtained with sufficient precision to constrain the models. Therefore we have compared ratios of different elements. Elemental ratios are more complicated than isotopic ratios since atomic fractionation effects may be important. The open symbols in Figure 2 correspond to the ratios after adjustment for volatility and mass fractionation effects (Meyer et al. 1997; see Binns et al. 2005 for details). The $^{12}$C/$^{16}$O ratio was not adjusted since the fraction of interstellar carbon in the solid state is poorly known. These adjusted ratios show improved agreement with the models, but the adjustments are not highly quantitative, and should be regarded as approximate values showing that ratios previously thought to be inconsistent with solar-system abundances may be consistent if GCRs are fractionated correctly on the basis of volatility and mass. (See Binns et al. 2005 for additional discussion.)

After adjustments for elemental fractionation, these data show an isotopic composition similar to that obtained by mixing ~20% of WR wind material with ~80% of material with solar-system composition. The largest enhancements with respect to solar-system ratios predicted by the WR models $^{12}C/^{16}O$, $^{22}Ne/^{20}Ne$, and $^{58}Fe/^{56}Fe$, are consistent with our observations. We take this agreement as evidence that WR star ejecta are very likely an important component of the cosmic-ray source material.

## 5. Discussion

Two independent approaches at modeling the WR contribution to GCRs (Higdon & Lingenfelter, 2005, and Binns et al. 2005) have shown that to explain the cosmic-ray data approximately 20% of the source material must be WR star ejecta. For WR material to be such a large component of the GCR source material, large quantities of it must be efficiently injected into the accelerator of GCRs. We believe that this is an important constraint for models of the origin of GCRs.

(There is no inconsistency between the ~20% estimates described above and the earlier ~2% quoted by Cassé & Paul (1982). Although the details of the Cassé & Paul calculations are not included in that paper, the 2% that they quote is material ejected only in the WC phase. In both the Higdon & Lingenfelter (2003) and the Meynet & Maeder (2003 & 2005) models, the ~20% of material includes all material ejected from the birth of the star to the end of the WC phase. Additionally, Cassé & Paul used the solar energetic particle ratio of 0.13 (Mewaldt, et al. 1979) instead of the solar wind value of 0.073 (Geiss, 1973) for $^{22}Ne/^{20}Ne$ to represent the solar system abundances in their estimate. When adjustments for these two factors are made, the fraction of WR material required in the Cassé & Paul calculations is ~20%, in good agreement with the more recent calculations.)

Another important constraint for the origin of cosmic rays, previously obtained from CRIS results, is the requirement that nuclei synthesized and accelerated by SNe are accelerated at least $10^5$ yr after synthesis. Wiedenbeck, et al. (1999) have previously shown that the $^{59}Ni$, which decays only by electron-capture, has completely decayed, within the measurement uncertainties, to $^{59}Co$ (Wiedenbeck et al. 1999). The $^{59}Ni$ can decay if it forms as dust grains or as gas in atomic or molecular form, or it could decay in a plasma environment. In the Meyer et al. (1997) and Ellison et al. (1997) model, it is assumed that it is initially accelerated as dust grains, since it is a refractory element. Dust grains are known to form in SN ejecta (e.g. SN1987A (Dwek, 1998) and Cas-A (Dunne et al. 2003)). It therefore appears likely that before acceleration, the $^{59}$Ni resides in dust grains where it decays and its $^{59}$Co daughter is later accelerated by SN shocks.

The average time between SN events within superbubbles is ~$3 \times 10^5$ years, depending upon the number of massive stars in the OB association (Higdon, et al. 1998). In Binns et al. (2005) we stated that this gives sufficient time between events for the $^{59}$Ni, which is synthesized in the SN explosions, to decay before its daughter product, $^{59}$Co, is accelerated to cosmic ray energies. However, Prantzos (2005) has suggested that WR winds could accelerate the newly synthesized nuclei on time scales short compared to the mean time between SNe, based on arguments contained in Parizot et al. (2004). The kinetic energy in WR winds is of the same order as is dissipated in supernova explosions (e.g. Lietherer et al. 1992). Prantzos argued that the superbubble environment experiences shock passages on times scales significantly shorter than the mean time between SN. He therefore suggested that the mean time between SNe is not the time scale that is relevant for $^{59}$Ni decay in the superbubble environment.

Most GCRs detected at Earth are believed to originate within less than 1kpc from the Sun (Ptuskin & Soutoul 1998). OB associations within 1 kpc of the Sun are composed of a few to as many as ~320 OB stars in SCO OB2, which is located ~140 pc from the Sun (de Zeeuw et al. 1999). Some OB associations are composed of stars that form at approximately the same time, i.e. they are coeval. For example, Per OB2 contains 17 OB stars, is located about 400 pc from the Sun (de Zeeuw et al. 1999), and has an age of ~3 MY (Blaauw, 1964). Some of the larger associations are composed of two or more subgroups, with the stars in each subgroup forming at about the same time, but with the subgroups themselves having differing ages. An example of this is the Orion OB1 association, which contains ~70 OB stars, with subgroups a, b, c, and d having ages of ~12, 7, 3, and 1 MY respectively (de Zeeuw et al. 1999). The mean time for subgroup formation, averaged over many associations, is ~4MY (Sargent, 1979). Subgroup formation is believed to result when a SN shock from a young OB association propagates into the molecular cloud in which the association is embedded, causing nearby massive star formation. The most massive stars have the shortest lifetimes; stars with initial mass greater than ~30$M_\odot$ have lifetimes of ≤6MY (Schaller, et al. 1992).

For the sake of argument, we will initially consider an OB association in which the stars are coeval. In Figure 3 we show the history of such an association.

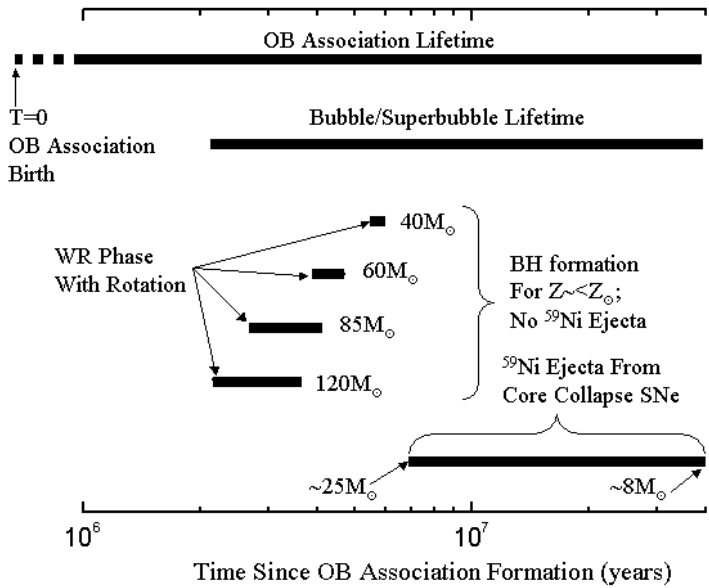

Figure 3—Diagram of the time evolution of a hypothetical OB association (see text).

Its lifetime begins with the localized condensation of molecular cloud material into massive stars at T=0 and ends when the least massive star that can undergo core-collapse (~8 $M_\odot$) ends its life as a supernova, ~40 MY later. Shortly after star formation, the most massive stars evolve into the Wolf-Rayet phase. Their high-velocity winds (~2,000-3,000 km/s) produce large low-density bubbles in the molecular cloud. The expanding shocks produced by the stars that undergo SN explosions coalesce and produce a superbubble. We show the time duration that the most massive stars spend in the WR phase, and the epoch for which that occurs in the OB association, for rotating stars with initial masses ranging from 40 to 120 $M_\odot$. We see that the most massive star modeled enters the WR phase roughly 2 MY after association birth, and the least massive star that can evolve into a WR star exits that phase roughly 4MY later. The exact low-initial-mass cutoff for entering the WR phase is model dependent and is believed to be between 25 $M_\odot$ and 40 $M_\odot$. (For details of these models of rotating stars, see Meynet & Maeder, 2003 & 2005; for the associated nucleosynthesis see Arnould et al. 2006.) The end of the WR phase of each star is followed by core collapse. So there is, at most, an ~4 MY interval in the life of a coeval OB association (~10% of its lifetime) for which acceleration of superbubble material by WR winds could occur. It is important to note that the most massive stars are very rare. The initial mass function for OB associations is not universally agreed upon, but is often taken to go approximately as $dN/dM \propto M^{-2.35}$ (Salpeter 1955; Higdon & Lingenfelter 2005). Therefore, in most OB associations, the most massive stars are not present, and the WR epoch is less than 4MY.

Let us suppose, as argued by Heger et al. (2003), that most stars with initial mass ≥ 40 $M_\odot$ and metallicity roughly solar or less do not undergo a SN explosion after core collapse, but instead "directly" form a black hole. Additionally, in their model, stars with initial mass 25-40 $M_\odot$ and metallicity roughly solar or less undergo core-collapse to form a black hole by "fallback", which results in a very weak SN shock with little ejecta. (We note that their model does not include the effects of rotation, which could change some of the details of their model predictions.) Daflon and Cunha (2004) have measured the metallicity of young OB stars in associations as a function of galacto-centric radius. Their results show that for OB associations within 1kpc of the Sun, most have metallicity that is solar or less. In the Heger et al. picture, stars with metallicity higher than roughly solar that undergo core collapse create supernovae of type SNIb,c. These supernovae are believed to result from WR stars, which have no hydrogen envelope, and thus no H emission. Additionally, there are massive stars that core-collapse into "hypernova", which are poorly understood, and estimated to occur in ~1-10% of the massive core-collapse events (Fryer et al. 2006). To the extent that this is a correct picture we see that a substantial fraction of core-collapse events during the WR epoch will not eject large amounts of newly synthesized material, including $^{59}$Ni, into the superbubble. Thus the predominant material that is available for acceleration by the WR winds appears to be wind material ejected from the association stars since their birth, plus any normal ISM that is in the vicinity.

Looking again at Figure 3, we see that those stars with mass low enough so that they do not enter the WR phase (~8 $M_\odot$ ≤ M ≤ 25 $M_\odot$) undergo core-collapse as SNe in which $^{59}$Ni is synthesized and injected into the superbubble. The most massive of these stars will undergo SN explosions first with subsequent SNe accelerating the material previously injected into the superbubble.

In this simple picture it appears that the injection of the $^{22}$Ne-rich wind material from WR stars and the injection of $^{59}$Ni from the SNe of stars with initial mass 8 $M_\odot$ ≤ M ≤25 $M_\odot$ are largely separated in time. Thus the appropriate time scale for acceleration of most SN ejecta would be the time between SN shocks after the WR epoch in superbubbles, not the shorter time scales associated with WR shocks in the WR epoch. The SN rate depends upon the number of stars in the OB association and has a time dependence related to the mass distribution of stars in the association. Since the time between SNe is typically ~3 × $10^5$ years for a large association (Higdon & Lingenfelter, 1998), and the $^{59}$Ni halflife for decay is 7.5 x $10^4$ years, in this picture, there is sufficient time for it to decay to $^{59}$Co.

For the fraction of OB associations that are composed of subgroups with differing ages, this simple picture needs to be modified since the WR winds from younger subgroups occur during the time period when substantial $^{59}$Ni is being ejected by SNe in older subgroups. However the fraction of the superbubble lifetime for which WR winds coexist with SN ejecta is still relatively small owing to the brief WR epoch. This is particularly true when you consider that there are many more WR stars at the light end of the mass spectrum than at the heavy end (Higdon & Lingenfelter, 2005), so the WR epoch in most subgroups is substantially shorter than 4MY. Thus, in this picture, the fraction of $^{59}$Ni that could be accelerated by WR winds, summing over all superbubbles in our neighborhood, is still relatively small.

It must be acknowledged that the superbubble environment is very complex. In addition to shocks from WR winds and SNe shocks, the winds and mass loss of OB stars in phases other than WR are very significant and may produce substantial shocks, though not as significant as either WR or SNe shocks (Parizot, 2004; Bykov, 2001). Furthermore, although the average time between SN events in an OB association is long compared to the $^{59}$Ni half-life, there will be a fraction of SN events that occur on shorter time scales, thus resulting in acceleration of recently synthesized $^{59}$Ni before it can decay. So, although the picture presented above seems useful in understanding how a substantial fraction of $^{59}$Ni can decay, the actual situation is likely more complex.

Additionally, we have argued in Binns et al. (2006) that although WR winds do contain roughly the same amount of kinetic energy (~$10^{51}$ ergs) as supernova explosions, the power in the WR termination shocks is about a factor of 10 less than in SNRs, which have lifetimes of ~$10^4$ years (Higdon 2006). Furthermore, it is possible that the WR shocks would only accelerate $^{59}$Ni nuclei to relatively low energies, where they would be only partially stripped of their orbital electrons, and the $^{59}$Ni could still decay. In a power law spectrum that one gets from shock acceleration, most of the nuclei are at low energy where they can decay.

Thus it appears that the observation by Wiedenbeck et al. (1999) that all or most of the $^{59}$Ni in GCRs has decayed to $^{59}$Co is likely consistent with the OB-association origin of galactic cosmic rays. We conclude that the good agreement of our isotopic data with WR models suggests that OB associations are the likely source of at least a substantial fraction of GCRs.

# Acknowledgements

The authors wish to thank J.C. Higdon and N. Prantzos for helpful discussions. This research was supported in part by the National Aeronatutics and Space Administration at Caltech, Washington University, the Jet Propulsion Laboratory, and Goddard Space Flight Center (under Grants NAG5-6912 and NAG5-12929).